\documentclass[12pt,preprint]{aastex}
\shorttitle{GRB~050319 OT Observations}
\shortauthors{Quimby et al.}

\begin{document}

\title{Early-Time Observations of the GRB~050319 Optical Transient}

\author{
Quimby,~R.~M.\altaffilmark{1},
Rykoff,~E.~S.\altaffilmark{2}, 
Yost,~S.~A.\altaffilmark{2},
Aharonian,~F.\altaffilmark{3},
Akerlof,~C.~W.\altaffilmark{2},
Alatalo,~K.\altaffilmark{2,4},
Ashley,~M.~C.~B.\altaffilmark{5}, 
G\"o\u{g}\"u\c{s},~E.\altaffilmark{6},
G\"{u}ver, T.\altaffilmark{7},
Horns,~D.\altaffilmark{3},
Kehoe,~R.~L.\altaffilmark{8},
K{\i}z{\i}lo\v{g}lu,~\"{U}.\altaffilmark{9},
McKay,~T.~A.\altaffilmark{2},
\"{O}zel,~M.\altaffilmark{10},
Phillips,~A.\altaffilmark{5}, 
Schaefer,~B.~E.\altaffilmark{11},
Smith,~D.~A.\altaffilmark{12},
Swan,~H.~F.\altaffilmark{2},
Vestrand,~W.~T.\altaffilmark{13}, 
Wheeler,~J.~C.\altaffilmark{1},
Wren,~J.\altaffilmark{13}
}

\altaffiltext{1}{Department of Astronomy, University of Texas, Austin, TX
        78712, quimby@astro.as.utexas.edu, wheel@astro.as.utexas.edu}
\altaffiltext{2}{University of Michigan, 2477 Randall Laboratory, 450 Church
        St., Ann Arbor, MI, 48104, erykoff@umich.edu, akerlof@umich.edu,
        kalatalo@umich.edu, tamckay@umich.edu,
        hswan@umich.edu, sayost@umich.edu}
\altaffiltext{3}{Max-Planck-Institut f\"{u}r Kernphysik, Saupfercheckweg 1,
        69117 Heidelberg, Germany, Felix.Aharonian@mpi-hd.mpg.de,
        horns@mpi-hd.mpg.de}
\altaffiltext{4}{UC Berkeley Astronomy, 601 Campbell Hall, Berkeley, CA, 94720, kalatalo@berkeley.edu}
\altaffiltext{5}{School of Physics, Department of Astrophysics and Optics,
        University of New South Wales, Sydney, NSW 2052, Australia,
        mcba@phys.unsw.edu.au, a.phillips@unsw.edu.au}
\altaffiltext{6}{Sabanc{\i} University, Orhanl{\i}-Tuzla 34956 Istanbul,
        Turkey, ersing@sabanciuniv.edu}
\altaffiltext{7}{Istanbul University Science Faculty, Department of Astronomy
        and Space Sciences, 34119, University-Istanbul, Turkey, 
        tolga@istanbul.edu.tr}
\altaffiltext{8}{Dept. of Physics, Southern Methodist University, Dallas, TX
        75275, kehoe@physics.smu.edu}
\altaffiltext{9}{Middle East Technical University, 06531 Ankara, Turkey,
        umk@astroa.physics.metu.edu.tr}
\altaffiltext{10}{\c{C}anakkale Onsekiz Mart \"{U}niversitesi, Terzio\v{g}lu
        17020, \c{C}anakkale, Turkey, m.e.ozel@ibu.edu.tr}
\altaffiltext{11}{Department of Physics and Astronomy, Louisiana State
        University, Baton Rouge, LA 70803, schaefer@lsu.edu}
\altaffiltext{12}{Guilford College, 5800 W. Friendly Ave., Greensboro, NC, 27410, dsmith4@guilford.edu}
\altaffiltext{13}{Los Alamos National Laboratory, NIS-2 MS D436, Los Alamos, NM
        87545, vestrand@lanl.gov, jwren@nis.lanl.gov}

\begin{abstract}
We present the unfiltered ROTSE-III light curve of the optical
transient associated with GRB~050319 beginning 4~s after the cessation
of $\gamma$-ray activity. We fit a power-law function to the data
using the revised trigger time given by \citet{chincarini05}, and a
smoothly broken power-law to the data using the original trigger
disseminated through the GCN notices. Including the RAPTOR data from
\citet{wozniak2005}, the best fit power-law indices are $\alpha=-0.854
\pm 0.014$ for the single power-law and $\alpha_1=-0.364
^{+0.020}_{-0.019}$, $\alpha_2= -0.881^{+0.030}_{-0.031}$, with a
break at $t_b= 418^{+31}_{-30}$~s for the smoothly broken
fit. We discuss the fit results with emphasis placed on the importance
of knowing the true start time of the optical transient for this
multi-peaked burst. As {\it Swift} continues to provide prompt GRB
locations, it becomes more important to answer the question, ``when
does the afterglow begin'' to correctly interpret the light curves.

\end{abstract}

\section{Introduction}

The precise localization and prompt dissemination of gamma-ray bursts
(GRBs) from {\it Swift} has opened the very early time domain of GRB
afterglows to exploration. Prior to {\it Swift}, the few bursts with
early afterglow detections engendered an assumption that bright
optical flashes commonly dominated the early light curves; however,
the growing sample of {\it Swift} bursts shows, contrary to these
expectations, that this phenomenon is rare and in fact many early
light curves show a deficit in flux compared to a backward
extrapolation of late time behavior. GRB~050319 adds to the growing
sample of such bursts with early-time optical observations and defines
new challenges to interpretation.

The position of GRB~050319 ({\it Swift} trigger 111622) was
distributed as a Gamma-ray Burst Coordinates Network (GCN) notice on
2005 March 19 at 09:31:38 UT, with a $4\arcmin$ radius error box. The
notice was issued after a single fast rise exponential decay (FRED)
profile triggered the BAT at 09:31:18.44 UT (hereafter
$t_{trigger1}$); however, \citet{chincarini05} report that re-analysis
of the prompt BAT light curve reveals the burst actually began 137~s
earlier ($t_{trigger0}=t_{trigger1}-137$~s), but this occurred during
a slew so no alert was issued. \citet{cusumano2005} give the starting
time for GRB~050319 as 09:29:02.70. The BAT light curve thus consists
of several peaks with combined $T_{90}=149.6\pm0.7$~s and a 15-350 keV
fluence of $1.6\times10^{-6}\,\mathrm{erg}\,\mathrm{cm}^{-2}$. There
are two principle peaks in the $\gamma$-ray light curve; the last
peak alone had a $7.3\times10^{-7}\,\mathrm{erg}\,\mathrm{cm}^{-2}$
fluence and $T_{90}=23.5$~\citep{chincarini05}.

The XRT began observations of the afterglow at 9:32:45.53
UT~\citep{krimm2005}. Adopting $ t_0=t_{trigger0}$ as the starting
point for the afterglow, \citet{chincarini05} find an initial steep
decline in the X-ray light curve with a power-law index of
$\alpha_1=-7.64 \pm 3.83$, which breaks at 329~s to $\alpha_2=-0.50
\pm 0.08$, and a second break to $\alpha_3=-2.07 \pm 0.06$ that occurs
at $t_{trigger0}+20.5\,\mathrm{hrs}$, where all values are from their
first fitting method. \citet{cusumano2005} give the slopes as
$\alpha_1=-5.53\pm0.67$, $\alpha_2=-0.54\pm0.04$, $\alpha_3=-1.14\pm0.2$
with breaks at $384\pm22$~s and $7.2\pm1.9$~hrs, and suggest the
initial fast decay may represent the low energy tail from the prompt
emission.

\citet{fynbo2005} report strong absorption lines in the optical
spectra, indicating a redshift of $z=3.24$.

In this letter, we report on the early-time optical observations of
GRB~050319 with the ROTSE-IIIb (Robotic Optical Transient Search
Experiment) telescope located at McDonald Observatory, Texas. The
observations are described in \S{\ref{observations}} and the reduction
of the data is detailed in \S{\ref{analysis}}. In \S{\ref{results}} we
present power-law function fits to the light curves exploring both
$t_{trigger0}$ and $t_{trigger1}$ as the start of the optical
emission. We end in \S{\ref{discussion}} with a discussion of the
starting time and implications of the multi-peaked burst.

\section{Observations}\label{observations}

The ROTSE-III array is a worldwide network of four 0.45~m robotic,
automated telescopes, built for fast ($\sim 6$ s) responses to GRB
triggers from satellites such as {\it Swift}.  They have a wide ($1\fdg85
\times 1\fdg85$) field of view imaged onto a Marconi $2048\times2048$
back-illuminated thinned CCD, and operate without filters.  The
ROTSE-III systems are described in detail in \citet{akmrs03}.

ROTSE-IIIb responded automatically to the GCN notice in under 8~s with
the first exposure starting at 09:31:45.5 UT, just 4~s after the
cessation of $\gamma$-ray activity.  The automated scheduler began a
program of ten 5-s exposures, ten 20-s exposures, and 149 60-s
exposures before the burst position dropped below our elevation
limit. Strong winds introduced tracking errors, which degraded the
quality of the initial images. Near real-time analysis of the
ROTSE-III images identified a $16^{th}$ magnitude fading source at
$\alpha=10^h16^m47\fs9$, $\delta=+43\arcdeg32\arcmin54\farcs5$
(J2000.0) that was not on the Digitized Sky Survey red plates, which
we reported via the GCN Circulars within 25 minutes of the
burst~\citep{rsq05}. Scattered clouds began to reduce the transparency
starting 22 minutes into the response. After 84 minutes, the clouds
thickened and the remaining images are not usable.

\section{Analysis}\label{analysis}

The raw ROTSE-III images were processed by an automatic script to
eliminate the dark current, and were normalized using a flat field
constructed from twilight exposures. We then performed relative
photometry on the optical transient (OT) and nearby objects using
RPHOT, a custom, interactive program implemented in IDL and based
around the DAOPHOT routines~\citep{stetson87} ported to IDL by
\citet{landsman89}. RPHOT measures both circular aperture and PSF-fit
fluxes for objects and provides checks to determine which method
produces the best results based on the derived photometric precision
of field stars.

We first constructed a deep coadded frame to serve as a reference for
both the photometry and the astrometry. The OT is well detected
(S/N$>10$) on the reference image (REFIM). A set of fiducial reference
stars (REFSTARS) was chosen from the REFIM to identify nearby
($<12'$), isolated stars that were not flagged as either saturated nor
blended by SExtractor~\citep{bertin_arnouts1996}. An initial list is
generated automatically, and remaining sources that appear to deviate
from the stellar PSF are removed by hand. The 57 REFSTARS selected are
used to derive magnitude zeropoints relative to the REFIM.

To determine the sky value local to each object under consideration,
we calculate a Gaussian weighted asymmetric clipped mean of the pixels
in an annulus of $2r$ to $3r$, where $r$ is 3.5 pixels or the local
FWHM, which ever is larger. The calculation is iterative, and pixels
$5\sigma$ below the mean and the adjacent pixels are rejected, where
$\sigma$ is the current estimate of the Gaussian width. Initially,
pixels $3\sigma$ above the mean and their neighbors are also rejected,
and this clipping threshold is raised with each successive
iteration. In addition, the OT sky annulus is given special
consideration; using the deep REFIM, all detected sources in the OT
sky annulus are masked, and this mask is propagated to the other
images. When a pre-existing mask is used, clipping is still performed
but with higher initial tolerances. When a large fraction of the sky
annulus is masked, it is enlarged to ensure the formal error in the
local sky calculation remains low.

With the OT and reference stars selected and any sky mask set, RPHOT
steps through the images and displays the full active region, a
rectangular area which encompasses the OT and all of the REFSTARS, and
a close up of the OT and its sky annulus. We looked for any global or
local problems which might interfere with the photometry. As each
image is displayed, RPHOT matches the REFSTARS up to the REFIM using
the RA, DEC solution generated from objects identified by
SExtractor. The matched REFSTARS are then used to determine the
coordinate mapping from the REFIM. Outliers are rejected from the
final solution and later their positions are recalculated using the
final transformation. When few of the REFSTARS are detected on an
image due to a combination of short exposure times or poor weather,
the solution based on the full frame is used if the transformation
residuals are smaller. The median transformation residuals for these
data are typically $\sim 0.1$ pixels. This transformation is then used
to map the OT location as found on the REFIM to each image.

Once the OT and the reference stars have been located on a given
image, aperture photometry is performed in a series of concentric,
circular apertures ranging from 0.4 to 2.0 times the local FWHM of the
image. In addition, a 3.5 pixel ($11.4''$) fixed aperture was
used. The local sky values and standard deviations are set using the
sky annuli and weighting as described above. RPHOT calculates the
weighted average of flux ratios of the reference stars to the REFIM in
each aperture in order to derive the relative magnitude offsets.

We finally use the standard DAOPHOT routines to calculate the PSF-fit
fluxes. As the PSF does vary significantly across the detector, we
selected well detected objects on each image within $14'$ of the OT to
construct the PSF fitting template; this radius represents a balance
between improved template accuracy gained from an increase in the
number of objects included and deviations from the OT's PSF at larger
separations that degrade the template. We did not modify the DAOPHOT
routines to fix the object centroids for PSF fitting, but allowed them
to move as the fitting required, although measurements where the
centroid moved by more that 0.75 times the local FWHM were discarded.

The magnitude scale is set in an absolute sense by calibrating
reference stars on the REFIM to a given system. The ROTSE-III
telescopes operate without a filter, and the peak sensitivity falls in
the R band. We calibrated the magnitude scale using 14 REFSTARS with
R-band values determined by \citet{henden2005}, and denote these
magnitudes as $C_{R}$. The stars used for calibration have colors in
the range $0.3 < V-R < 1.0$, with a median of 0.54. Our $C_{R}$
magnitudes may differ from the R band values if the spectral energy
distribution of the OT differs from that of the median reference
star. We can estimate this offset by adopting a blackbody with an
effective temperature of 5560 K for our median reference star, and by
assuming that the OT spectrum can be represented by $F_{\nu}\propto
\nu^{\beta}$ with $-1.5 < \beta < -0.5$. With these assumptions, the
unfiltered to R band flux ratio is greater for the OT than for the
median reference star, which makes the $C_{R}$ values 0.1 to 0.2 mag
brighter than the true R band magnitudes. The correction factor,
however, is sensitive to the amount of absorption along the line of
sight as the measured redshift for the OT ($z=3.24$) places rest-frame
Lyman-$\alpha$ within our bandpass. Depending on the amount of
absorption, this could then make the $C_{R}$ values up to 0.3 mag {\it
fainter} than the true R band magnitudes. Further note we have not
corrected our $C_{R}$ magnitudes for extinction.

After all the images are processed, RPHOT includes tools to assess the
data quality and check the consistency of the relative photometry via
the REFSTARS. The reference stars were found to all have flat light
curves and no trends were found for objects near the OT.

RPHOT initially displays only the S/N for the OT. This is to allow for
a co-addition scheme for the later data when the OT has faded near or
below the individual image limiting magnitudes to be investigated
without biasing the shape of the light curve. As OTs typically decay
as power-laws at later times, co-adding the frames into logarithmically
longer and longer time bins is generally needed to maintain the same
S/N. For the circular aperture photometry, actual co-addition of the
frames is not required, but rather a weighted average of the fluxes
can be used. The PSF-fitting, however, fails for weak or
non-detections and thus must be performed on coadded frames where the
S/N is above ~3.

Using the magnitude RMS and $\chi^2$ fits to the reference objects, we
determined that the PSF fitting produces the best results. Images
where the PSF fitting failed were coadded to bring the OT S/N above
3. To determine how to group the images for co-adding, we calculated
the average fluxes of the OT in the 0.72 FWHM aperture in sets of
images weighted by the flux error, which is effectively just the sky
noise because of the weak OT signal. We continued to add successive
images to a set until the S/N was greater than 3, and then we co-added
each set of images, again using the flux errors as weights.

\section{Results}\label{results}

\begin{figure}
\epsscale{1.0}
\plotone{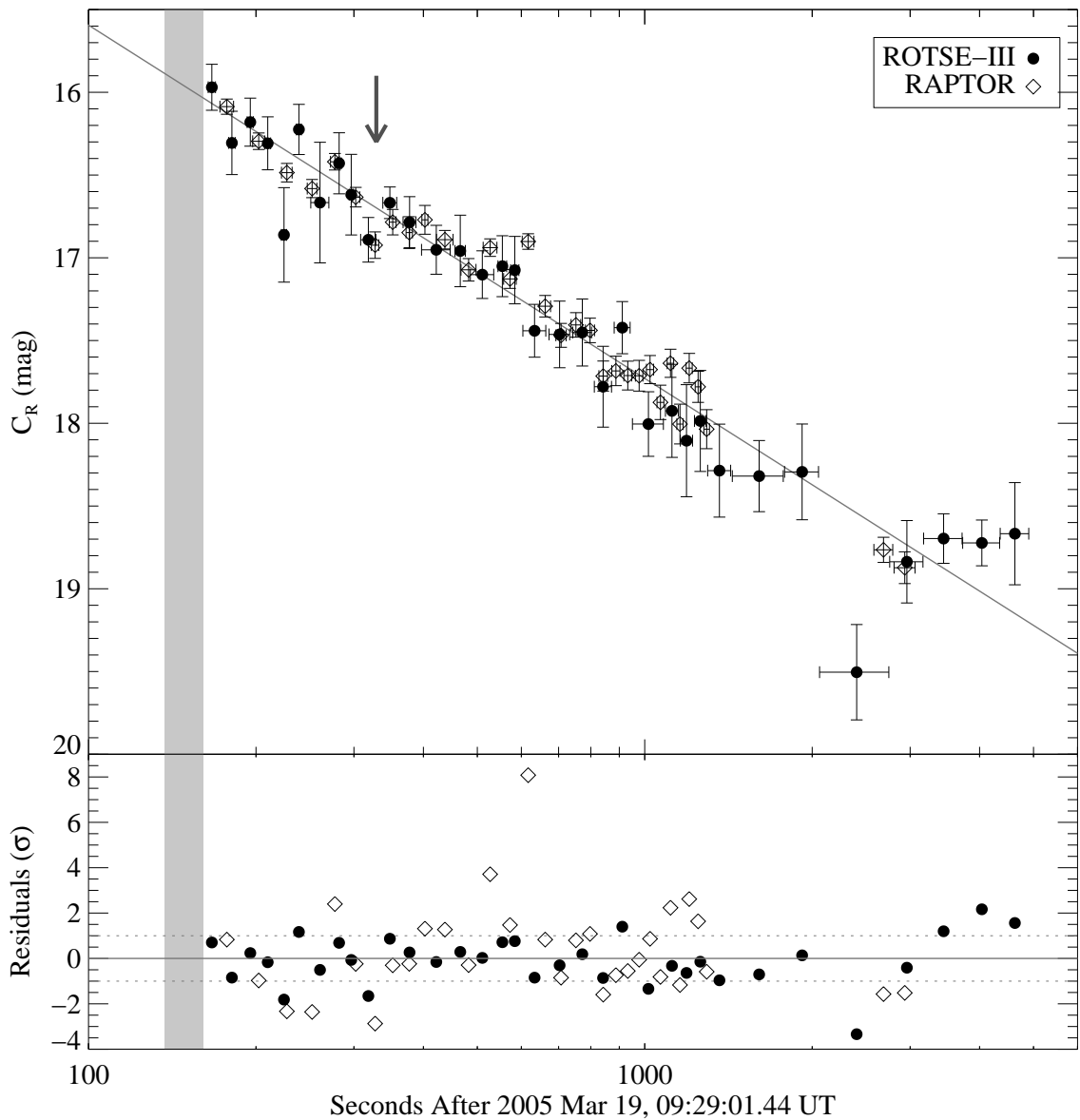}
\caption{ROTSE-III and RAPTOR light curve for the GRB~050319 optical
transient for $t_0=t_{trigger0}$. The line gives the best fit single
power-law to the combined data set, $\alpha=-0.854$. The vertical
shaded band marks the last, FRED-like peak in the $\gamma$-ray light
curve and the arrow marks the first break in the XRT light
curve. RAPTOR data have been shifted by 0.21 mag as described in the
text.}
\label{lc0}
\end{figure}

The PSF-fit magnitudes are listed in Table \ref{otmags} and shown in
Figure \ref{lc0} with times relative to $t_{trigger0}$
(ie. $t=t_{obs}-t_0$ where $t_{obs}$ is the time of the observation
and $t_0=t_{trigger0}$). We fit a function of the form
$f(t)=f_0t^{\alpha}$ to the data and find a power-law decline of
$\alpha=-0.894^{+0.034}_{-0.033}$, where we have integrated the
probability surface, $P(f_0,\alpha)$, over all $f_0$ and found the
most probable $\alpha$ and the corresponding interval containing 68\%
of the total probability. The reduced $\chi^2$ for the best fit is
$\chi^2/\rm{DoF}=52.4/(34-1)=1.59$. A single point at
$t_{obs}-t_{trigger0}\sim 2400\,\rm{s}$ alone contributes 19.4 to the
$\chi^2$. We checked the PSF-fit magnitudes with the aperture
magnitudes for this point and found similar results. Further, we
inspected the photometry for neighboring objects and did not find any
anomalous behavior, although there was a spike in cloud opacity during
the effective integration.

Also shown in Figure \ref{lc0} are the unfiltered RAPTOR data for the
OT~\citep{wozniak2005}. We have subtracted 0.21 mag, which represents
the systematic zeropoint offset we find between the \citet{henden2005}
field calibration and the USNO-B1.0 R-band magnitudes for our data,
from the RAPTOR data\footnote{RAPTOR magnitudes from
\citet{wozniak2005} were calibrated using the USNO-B1.0 R2 magnitudes;
their reference to USNO-A2.0 R2 magnitudes is a misprint (Vestrand
2005, priv. com.)}. This zeropoint shift appears to fully account for
the systematic discrepancy between the ROTSE-III and RAPTOR
magnitudes and we assume the filter responses are close enough that
introduction of a color term is not necessary. The best fit power-law
index for the combined ROTSE-III and RAPTOR data is $\alpha=-0.854 \pm
0.014$, with a reduced $\chi^2/\rm{DoF}$ of $178.1/(66-1)=2.74$. The
ROTSE-III point at $t_{obs}-t_{trigger0}\sim 2400\,\rm{s}$ contributes
28.8 to this $\chi^2$, and a single RAPTOR point at
$t_{obs}-t_{trigger0}\sim 600\,\rm{s}$ adds an additional 46.5. If we
remove the three data points differing by more than $3\sigma$ from the
above fit, the best fit value becomes $\alpha=-0.844 \pm 0.015$ with
$\chi^2/\rm{DoF}=1.459$

\begin{figure}
\epsscale{1.0}
\plotone{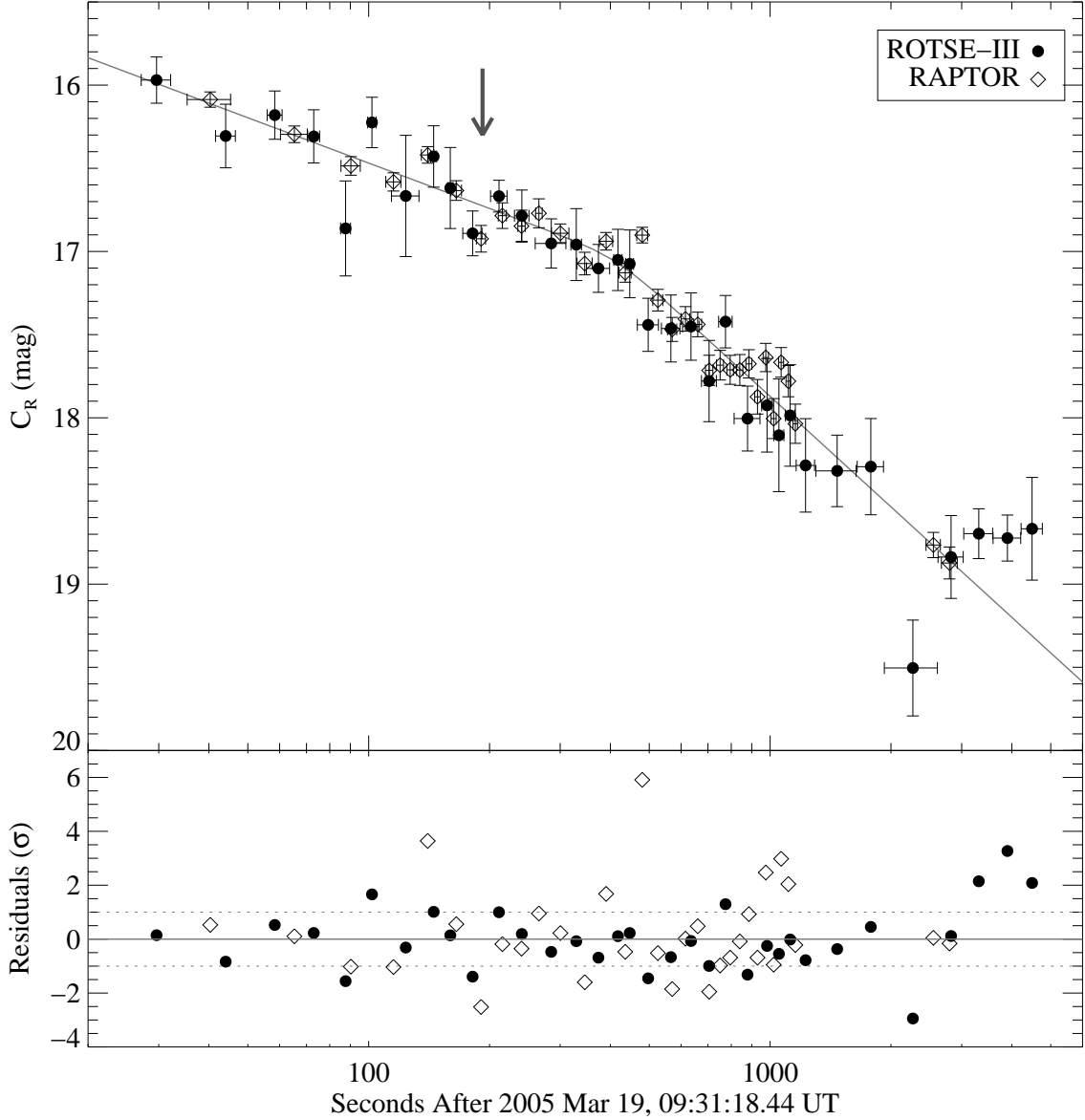}
\caption{ROTSE-III and RAPTOR light curve for the GRB~050319 optical
transient for $t_0=t_{trigger1}$.  The line gives the best smoothly
broken power-law fit to the joint data set with $\alpha_1=-0.36$,
$\alpha_2=-0.88$, and $t_b=418$~s. The smoothness parameter was fixed
at $s=20$. The arrow marks the first break in the XRT light curve.}
\label{lc1}
\end{figure}

Using the revised BAT trigger time, $t_{trigger0}$,
\citet{chincarini05} find a dramatic initial decline in the XRT light
curve (LC), which breaks to a slower decline at 329~s. This behavior
is not mirrored in the optical, nor are there indications of any
differences in the ROTSE-III and RAPTOR data before and after the
X-ray break; the optical LC simply continues the single power-law
decline. With $t_0=t_{trigger0}$, the initial X-ray decline is also
the steepest for any of the growing number of {\it Swift}
GRBs. However, \citet{chincarini05} find the X-ray LC is quite similar
to that of GRB~050318 if the start time coincides with the later BAT
trigger, $t_{trigger1}$. With this convention, the initial decline
becomes more shallow and is more typical of other XRT afterglows (for
example, see Figure 1 from \citealt{tagliaferri2005}). We therefore
investigated the impact of this change in epoch on the optical light
curve. Figure \ref{lc1} shows the ROTSE-III and RAPTOR data deviate
from a simple power-law decline with this choice for the afterglow
start time. \citet{wozniak2005} have analyzed the RAPTOR data using
$t_0=t_{trigger1}$ and found a broken power-law model gives a more
acceptable fit. Fitting a smoothly broken power-law of the form
\begin{equation}
f(t) = f_b 2^{1/s} [ (t/t_b)^{-s \alpha_1} + (t/t_b)^{-s \alpha_2} ]^{-1/s}\label{sbpl}
\end{equation}
\noindent with the smoothing parameter fixed at $s=20$ for a sharp
slope transition, the ROTSE-III data gives $\alpha_1=-0.354
^{+0.071}_{-0.062}$, $\alpha_2= -0.788^{+0.054}_{-0.060}$, and a break
time $t_b= 281^{+91}_{-69}$~s, with a best fit
$\chi^2/\rm{DoF}=1.57$. A joint fit to the ROTSE-III and RAPTOR data
gives $\alpha_1=-0.364 ^{+0.020}_{-0.019}$, $\alpha_2=
-0.881^{+0.030}_{-0.031}$, and a break time $t_b=
418^{+31}_{-30}$~s, with a best fit
$\chi^2/\rm{DoF}=2.24$. Again, the ROTSE-III and RAPTOR outliers
mentioned above add $19.9$ and $27.2$ to the $\chi^2$,
respectively. Removing the three $>3\sigma$ outliers improves the
joint fit to $\alpha_1=-0.367 \pm 0.022$, $\alpha_2= -0.864 \pm
0.034$, and a break time $t_b= 405 \pm 40$~s, with a best fit
$\chi^2/\rm{DoF}=1.51$.

It is difficult to explain such an early break in the optical LC with
the derived decay slopes in the context of the fireball model. One
possibility is that energy injection from a long lived inner engine
may be sustaining the optical emission until the break. If so, and
assuming the OT would otherwise have faded as a simple
power-law with $\alpha=\alpha_2$ from the first ROTSE-III point, then
the energy injected from $t_{obs}-t_{trigger1}=30$~s to $t_{b}$
increased the post-break optical flux by 3.9 times.

We have also considered the possibility that the break is due to a
synchrotron break. For example, if we naively assume the pre-break
data were taken around when the typical electron synchrotron
frequency, $\nu_m$, drops below our observing range then the observed
$\alpha_1$ could be the typical $-1/4$ index for fast cooling and a
constant-density ISM diluted by the transition to the steeper
decline. However, for the later decline we should have
$\alpha_2=-(3p-2)/4$, which results in, $p=1.84$, an unusual value for
the electron power-law index (note when $p<2$, there is an imposed
maximum in the distribution of electron Lorentz factors, which will
alter the relation of $p$ and $\alpha$). There is not a fixed value
for $p$ that predicts both $\alpha_1$ and $\alpha_2$ using the
relations for slow cooling with $1<p<2$ derived in
\citet{dai_cheng2001}. Other synchrotron breaks, such as the cooling
break, result in similarly atypical values for $p$. Further, the X-ray
LC breaks to a more shallow decline instead of the switch to a steeper
decline found in the optical at close to the same time, which is not
commensurate with a synchrotron break. The break is also not explained
by a jet break since it is not achromatic.

\section{Discussion}\label{discussion}

The afterglow of GRB050319 may have began at the first of two strong
$\gamma$-ray peaks with the steepest decline in the X-ray of any of
the GRBs captured by {\it Swift} and showed no correlated behavior in
its optical LC, or perhaps it began at the last peak in the
$\gamma$-rays and had a normal X-ray LC and an optical LC with a break
around 300~s between unusual power-law decline indices. In either case
there is no correlation between the X-ray and optical light curves:
the X-ray LC breaks do not coincide with optical breaks and the X-ray
decline rates do not match the optical slopes. The decline mismatch
means that the X-ray to optical color is continually changing and the
spectral slope $\beta$ (where $f_{\nu} \propto \nu^{\beta}$) is not a
constant from the X-ray to the optical.

In the context of the fireball model, reverse shock emission has been
predicted to dominate the early afterglow, giving a steep initial
decline~\citep{sari_piran1999}. Reverse shock emission has been used
to explain the initial rapid optical decline of the GRB~990123
afterglow, and may account for the steep decline in the early XRT
light curve for GRB~050319; however, this emission is expected to peak
in the optical or infrared and no such signal is observed in the
contemporaneous optical data. \citet{tagliaferri2005} and
\citet{kobayashi2005}, however, have discussed suppression of the
optical signal through inverse Compton scattering. In this case the
optical photons produced in the reverse shock are up-scattered, which
creates the fast decaying X-ray signal. \citet{kobayashi2005} show
that while inverse Compton effects can highly suppress optical
emission in the reverse shock, inverse Compton emission can be less
important in the forward shock. Up-scattering of the reverse shock
optical emission could help explain why there is no change in the
optical decay during the first break in the X-ray LC. However,
\citet{cusumano2005} suggest the initial fast X-ray decay may
represent the low energy tail from the prompt emission, in which case
no reverse shock emission was observed in any band.

The true behavior of the optical light curve is critically dependent
on the choice of $t_0$. In general, for accurate analysis the error in
$t_0$ must be much less than the epochs in which power-laws are to be
evaluated. Although the reduced $\chi^2$ is smaller for the smoothly
broken power-law fit to the combined ROTSE-III and RAPTOR data with
$t_0=t_{trigger1}$ than for the single power-law fit to the same data
with $t_0=t_{trigger0}$ (including the outliers in both cases),
neither model can acceptably account for the scatter shown in the
sample. If such fluctuations are intrinsic to early OT light curves,
it will be difficult to use statistical arguments to determine the
starting time, $t_0$, through fits to simple power-law functions. The
behavior predicted by the fits for the two choices of $t_0$ considered
differ by about 0.2 mag during later times with published magnitude
estimates, although the errors for the two models
overlap. \citet{wozniak2005} have shown using data from the GCN
Circulars that the decay rate of the OT appears to slow after 1.3
hours, and as a result we cannot use predictions for the late time
behavior from the single and smoothly broken power-law fits of the
early time data to constrain $t_0$.

If the optical emission began during the first peak in the
$\gamma$-rays then there are no deviations from a simple power-law in
the early phases that can be attributed to the last $\gamma$-ray
peak, even though the ROTSE-III data begin 27~s after the last peak
and just 4~s after the cessation of $\gamma$-ray activity. Therefore
this scenario leads to a physical difference in the first $\gamma$-ray
peak, which is followed by long lived optical emission, and the last
$\gamma$-ray peak, which has no detected optical emission and at most
optical emission several times fainter than that associated with the
first peak.

As the GRB050319 OT clearly illustrates, shifting $t_0$ to a later
time can turn a simple power-law into an apparent broken power
law. Because this shift, $t'_0 = t_0+\Delta t$, makes the logarithmic
difference between two epochs larger while the drop in flux remains
the same, the early light curve ($t'<\Delta t$) appears to flatten out
and we infer $\alpha'_1 \approx 0$. For $t' \gg \Delta t$ the effect
is negligible and we have $\alpha'_2 \approx \alpha$. Setting
$\alpha_1=0$, $\alpha_1=\alpha$ and $s=-1/\alpha$ into equation
\ref{sbpl} gives $f(t) = f'_{0}(t+t_b)^\alpha$, which is identical to
a single power-law shifted by $t_b$. However, a broken power-law with
a sharp transition ($s \gg -1/\alpha_{2}$) can be distinguished from a
single power-law with an incorrect $t_0$ if the light curve is well
sampled and the error bars are $\sim 0.1$ mag or smaller, which
ROTSE-III can deliver. It is therefore possible at least to determine
if an apparent early light curve break is due to an error in the
adopted $t_0$ even for $\alpha_1 \approx 0$ if the transition is
sharp. There are events which exhibit just such behavior, such as the
optical transient to GRB~050801 \citep{rykoff2005}.

It is important to note that gamma-ray burst triggers are defined by
instrument response and software algorithms and do not necessarily
mark the start of the burst itself, much less the afterglow. There are
many examples where the $\gamma$-ray emission is detected prior to the
formal trigger. For example, GRB~050915B showed $\gamma$-ray activity
10~s prior to the formal trigger~\citep{falcone2005}, while emission
began at least 8s before the GRB~050908 trigger~\citep{sato2005} and
GRB~050827 started 15s before its
trigger~\citep{sakamoto2005}. \citet{lazzati2005} searched BATSE data
in the range $-200\rm{s} < t < t_{trigger}$ and found that about 20\%
of bursts showed evidence for precursor activity. Setting $t_0$ to the
precursor time would effectively steepen the observed afterglow
decline rate and could thus lead to a different interpretation of the
LC, such as the presence of reverse shock emission. There have also
been bright bursts like GRB~990123 that began with $\sim 15$s of weak
$\gamma$-ray emission and later showed bright
peaks~\citep{briggs1999}. Assuming a similar light curve behavior for
weaker bursts, the trigger time could be delayed in some cases. The
derived decline rate based on the trigger would then be slower than
the decline rate based on the true start of the burst, and the shift
could introduce an apparent break in the observed afterglow LC.

The extended and highly variable nature of long-duration GRBs suggests
the afterglow itself may not begin cleanly from a given epoch, but
rather we might expect a turn-on phase where energy injection drives
the optical emission and perhaps produces a highly variable light
curve similar to that of GRB~050319 with $t_0=t_{trigger1}$. However
if $t_{trigger0}$ did mark the start of the afterglow, then the lack
of a bulk departure from the simple power-law decline as a result of
energy injection related to the last $\gamma$-ray peak argues against
a $\gamma$-ray/optical correlation. As {\it Swift} continues to
provide prompt GRB locations, it becomes more important to answer the
question, ``when does the afterglow begin'' to correctly interpret the
light curves.

\acknowledgements

This work has been supported by NASA grants NNG-04WC41G and F006794, NSF grants
AST-0119685 and 0105221, the Australian Research Council, the University of New
South Wales, and the University of Michigan.  Work performed at LANL is
supported through internal LDRD funding.  Special thanks to the observatory
staff at McDonald Observatory, especially David Doss.

\bibliographystyle{aa}  

\begin{thebibliography}{19}
\expandafter\ifx\csname natexlab\endcsname\relax\def\natexlab#1{#1}\fi

\bibitem[{{Akerlof} {et~al.}(2003){Akerlof}, {Kehoe}, {McKay}, {Rykoff},
  {Smith}, {Casperson}, {McGowan}, {Vestrand}, {Wozniak}, {Wren}, {Ashley},
  {Phillips}, {Marshall}, {Epps}, \& {Schier}}]{akmrs03}
{Akerlof}, C.~W., {Kehoe}, R.~L., {McKay}, T.~A., {et~al.} 2003, \pasp, 115,
  132

\bibitem[{{Bertin} \& {Arnouts}(1996)}]{bertin_arnouts1996}
{Bertin}, E. \& {Arnouts}, S. 1996, \aaps, 117, 393

\bibitem[{{Briggs} {et~al.}(1999){Briggs}, {Band}, {Kippen}, {Preece},
  {Kouveliotou}, {van Paradijs}, {Share}, {Murphy}, {Matz}, {Connors},
  {Winkler}, {McConnell}, {Ryan}, {Williams}, {Young}, {Dingus}, {Catelli}, \&
  {Wijers}}]{briggs1999}
{Briggs}, M.~S., {Band}, D.~L., {Kippen}, R.~M., {et~al.} 1999, \apj, 524, 82

\bibitem[{{Chincarini} {et~al.}(2005){Chincarini}, {Moretti}, {Romano},
  {Covino}, {Tagliaferri}, {Campana}, {Goad}, {Kobayashi}, {Zhang}, {Angelini},
  {Banat}, {Barthelmy}, {Beardmore}, {Boyd}, {Breeveld}, {Burrows}, {Capalbi},
  {Chester}, {Cusumano}, {Fenimore}, {Gehrels}, {Giommi}, {Hill}, {Hinshaw},
  {Holland}, {Kennea}, {Krimm}, {La Parola}, {Mangano}, {Marshall}, {Mason},
  {Nousek}, {O'Brien}, {Osborne}, {Perri}, {Meszaros}, {Roming}, {Sakamoto},
  {Schady}, {Still}, \& {Wells}}]{chincarini05}
{Chincarini}, G., {Moretti}, A., {Romano}, P., {et~al.} 2005, astro-ph/0506453

\bibitem[{{Cusumano} {et~al.}(2005){Cusumano}, {Mangano}, {Angelini},
  {Barthelmy}, {Beardmore}, {Burrows}, {Campana}, {Cannizzo}, {Capalbi},
  {Chincarini}, {Gehrels}, {Giommi}, {Goad}, {Hill}, {Kennea}, {Kobayashi}, {La
  Parola}, {Malesani}, {Meszaros}, {Mineo}, {Moretti}, {A. Nousek}, {O'Brien},
  {Osborne}, {Pagani}, {Page}, {Perri}, {Romano}, {Tagliaferri}, \&
  {Zhang}}]{cusumano2005}
{Cusumano}, G., {Mangano}, V., {Angelini}, L., {et~al.} 2005, \apj\,accepted
  (astro-ph/0509689)

\bibitem[Dai \& Cheng(2001)]{dai_cheng2001} Dai, Z.~G., \& Cheng, 
K.~S.\ 2001, \apjl, 558, L109 

\bibitem[{{Falcone} {et~al.}(2005){Falcone}, {Beardmore}, {Burrows},
  {Cummings}, {Fox}, {Gehrels}, {Holland}, {Kennea}, {Krimm}, {Pagani},
  {Palmer}, {Parsons}, {Rol}, {Roming}, \& {Shady}}]{falcone2005}
{Falcone}, A., {Beardmore}, A., {Burrows}, D., {et~al.} 2005, {GCN Circular}
  3987

\bibitem[{{Fynbo} {et~al.}(2005){Fynbo}, {Hjorth}, {Jensen}, {Jakobsson},
  {Moller}, \& {N{\"a}r{\"a}nen}}]{fynbo2005}
{Fynbo}, J.~P.~U., {Hjorth}, J., {Jensen}, B.~L., {et~al.} 2005, {GCN Circular}
  3136

\bibitem[{{Henden}(2005)}]{henden2005}
{Henden}, A. 2005, {GCN Circular} 3454

\bibitem[{{Kobayashi} {et~al.}(2005)}]{kobayashi2005}
{Kobayashi}, S., {Zhang}, B., {Meszaros}, P., {et~al.} 2005, astro-ph/0506157

\bibitem[{{Krimm} {et~al.}(2005){Krimm}, {Still}, {Barthelmy}, {Barbier},
  {Campana}, {Capalbi}, {Chester}, {Cummings}, {Fenimore}, {Gehrels}, {Goad},
  {Godet}, {Greiner}, {Grupe}, {Hullinger}, {Parola}, {Mangano}, {Markwardt},
  {Meszaros}, {Morris}, {Nousek}, {Page}, {Palmer}, {Parsons}, {Sakamoto},
  {Sato}, {Suzuki}, {Tagliaferri}, \& {Tueller}}]{krimm2005}
{Krimm}, H., {Still}, M., {Barthelmy}, S., {et~al.} 2005, {GCN Circular} 3117

\bibitem[{{Landsman}(1989)}]{landsman89}
{Landsman}, W.~B. 1989, \baas, 21, 784

\bibitem[{{Lazzati}(2005)}]{lazzati2005}
{Lazzati}, D. 2005, \mnras, 357, 722

\bibitem[{{Rykoff} {et~al.}(2005{\natexlab{a}}){Rykoff}, {Schaefer}, \&
  {Quimby}}]{rsq05}
{Rykoff}, E., {Schaefer}, B., \& {Quimby}, R. 2005{\natexlab{a}}, {GCN
  Circular} 3116

\bibitem[{{Rykoff} {et~al.}(2005{\natexlab{b}}){Rykoff}, {Magnano}, {Yost},
  {Aharonian}, {Akerlof}, {Burrows}, {Gehrels}, {G\"{u}ver}, {Horns}, {McKay},
  {\"{O}zel}, {Phillips}, {Quimby}, {Rujopakarn}, {Schaefer}, {Smith}, {Swan},
  {Wheeler}, {Wren}, \& {Yuan}}]{rykoff2005}
{Rykoff}, E.~S., {Magnano}, V., {Yost}, S.~A., {et~al.} 2005{\natexlab{b}},
  submitted

\bibitem[{{Sakamoto} {et~al.}(2005){Sakamoto}, {Barbier}, {Barthelmy},
  {Cummings}, {Fenimore}, {Gehrels}, {Hinshaw}, {Hullinger}, {Krimm},
  {Markwardt}, {McLean}, {Palmer}, {Parsons}, {Sato}, \&
  {Tueller}}]{sakamoto2005}
{Sakamoto}, T., {Barbier}, L., {Barthelmy}, S., {et~al.} 2005, {GCN Circular}
  3894

\bibitem[{{Sari} \& {Piran}(1999)}]{sari_piran1999}
{Sari}, R. \& {Piran}, T. 1999, \apj, 520, 641

\bibitem[{{Sato} {et~al.}(2005){Sato}, {Barbier}, {Barthelmy}, {Boyd},
  {Cummings}, {Hullinger}, {Fenimore}, {Gehrels}, {Krimm}, {Markwardt},
  {Palmer}, {Parsons}, {Sakamoto}, {Tueller}, \& {Voges}}]{sato2005}
{Sato}, G., {Barbier}, L., {Barthelmy}, S., {et~al.} 2005, {GCN Circular} 3951

\bibitem[{{Stetson}(1987)}]{stetson87}
{Stetson}, P.~B. 1987, \pasp, 99, 191

\bibitem[Tagliaferri et al.(2005)]{tagliaferri2005} Tagliaferri, G., et 
al.\ 2005, \nat, 436, 985 

\bibitem[{{Wo{\'z}niak} {et~al.}(2005){Wo{\'z}niak}, {Vestrand}, {Wren},
  {White}, {Evans}, \& {Casperson}}]{wozniak2005}
{Wo{\'z}niak}, P.~R., {Vestrand}, W.~T., {Wren}, J.~A., {et~al.} 2005, \apjl,
  627, L13

\end{thebibliography}

\begin{deluxetable}{rrcccc}
\tablewidth{0pt}
\tablecaption{}
\tabletypesize{\scriptsize}
\tablehead{
  \colhead{$t_{\mathrm{start}}$ (s)} &
  \colhead{$t_{\mathrm{end}}$ (s)} &
  \colhead{Exp. (s)} &
  \colhead{$C_R$} &
  \colhead{$\sigma$}
}
\startdata
    164.12 &     169.12 &     5 &    15.97 &    0.14 \\
    178.54 &     183.54 &     5 &    16.31 &    0.19 \\
    192.86 &     197.86 &     5 &    16.18 &    0.15 \\
    207.48 &     212.48 &     5 &    16.31 &    0.16 \\
    222.10 &     227.10 &     5 &    16.86 &    0.29 \\
    236.42 &     241.42 &     5 &    16.22 &    0.15 \\
    250.94 &     270.46 &    10 &    16.67 &    0.36 \\
    279.68 &     284.68 &     5 &    16.43 &    0.18 \\
    294.20 &     299.20 &     5 &    16.62 &    0.24 \\
    308.52 &     328.52 &    20 &    16.89 &    0.13 \\
    338.17 &     358.17 &    20 &    16.67 &    0.10 \\
    367.72 &     387.72 &    20 &    16.79 &    0.15 \\
    396.96 &     446.82 &    40 &    16.95 &    0.15 \\
    455.96 &     475.96 &    20 &    16.96 &    0.22 \\
    485.81 &     535.16 &    40 &    17.10 &    0.14 \\
    544.71 &     564.71 &    20 &    17.05 &    0.18 \\
    573.95 &     593.95 &    20 &    17.07 &    0.20 \\
    603.81 &     663.81 &    60 &    17.44 &    0.16 \\
    672.96 &     732.96 &    60 &    17.46 &    0.20 \\
    742.21 &     802.21 &    60 &    17.45 &    0.20 \\
    811.56 &     871.56 &    60 &    17.78 &    0.24 \\
    881.22 &     941.22 &    60 &    17.42 &    0.16 \\
    950.47 &    1080.23 &   120 &    18.00 &    0.20 \\
   1089.38 &    1149.38 &    60 &    17.93 &    0.28 \\
   1158.84 &    1218.84 &    60 &    18.11 &    0.34 \\
   1228.60 &    1288.60 &    60 &    17.99 &    0.31 \\
   1297.75 &    1427.11 &   120 &    18.29 &    0.28 \\
   1436.97 &    1774.48 &   300 &    18.32 &    0.21 \\
   1784.24 &    2052.92 &   240 &    18.29 &    0.29 \\
   2062.07 &    2746.86 &   600 &    19.50 &    0.29 \\
   2756.52 &    3163.60 &   360 &    18.84 &    0.25 \\
   3172.75 &    3719.54 &   480 &    18.70 &    0.15 \\
   3728.69 &    4345.86 &   420 &    18.72 &    0.14 \\
   4355.22 &    4902.01 &   480 &    18.67 &    0.31 \\
\enddata

\tablecomments{$t_{start}$ and $t_{end}$ give the time since
$t_{trigger0}$ (2005 Mar 19, 09:29:01.44 UT) in the observer
frame. Exp is the total open shutter time. $C_R$ is the unfiltered
magnitude calibrated against the R-Band magnitudes of
\citet{henden2005}.}

\label{otmags}
\end{deluxetable}

\clearpage

\clearpage

\end{document}